# Significance of Quality Metrics during Software Development Process

[1]Poornima. U. S., [2] Suma. V
[1]Program Manager, MCA Department,
Acharya Institute of Management and Sciences
[1,2]Research and Industry Incubation Centre,
Dayananda Sagar Institutions, Bangalore
[1]uspaims@gmail.com, [2]sumavdsce@gmail.com

**Abstract:** In recent years, Software has become an indispensable part of every segment from simple Office Automation to Space Technology and E-mail to E-commerce. The evolution in Software architecture is always an open issue for researchers to address complex systems with numerous domain-specific requirements. Success of a system is based on quality outcome of every stage of development with proper measuring techniques. Metrics are measures of Process, Product and People ($P^3$) who are involved in the development process, acts as quality indicators reflecting the maturity level of the company. Several process metrics has been defined and practiced to measure the software deliverables comprising of requirement analysis through maintenance. Metrics at each stage has its own significance to increase the quality of the milestones and hence the quality of end product. This paper highlights the significance of software quality metrics followed at major phases of software development namely requirement, design and implementation. This paper thereby aims to bring awareness towards existing metrics and leads towards enhancement of them in order to reflect continuous process improvement in the company for their sustainability in the market.

**1. Introduction:** Overall success of a project begins with proper understanding of the problem space, project planning and scheduling, development process, and expertise people contribution, SQA activities with right set of metrics, tool set and documentation. Metrics are measures of activities, people involved and the product under development, gives an insight on their quality to make overall process successful. Several metrics are defined and adapted in software industries, however, it has been an open issue having more scope to improvise $P^3$ to address problems with different levels of complexity. Since quality of process has an impact on quality of end product, process metrics plays significant role contributing to end product quality [7][8][9][10][11][12][13][14].Different process metrics, their significance and applicability have been a major open issue in the literature.

**2. Software Requirement Metrics:**

Requirements collection, analysis, specification and documentation are crucial steps of Software Development Life Cycle need more attention and regular refinement to keep the process under control. Metrics are defined for requirement specification to documentation as a part of Requirement Engineering to make the phase clear for further activities. Object Oriented System Design focuses on framing static (Class diagrams) and dynamic (Object interaction diagrams) models as a part of requirements collection. Use case diagrams are used to gather set of scenarios and overall behaviour of the system in user's perception. The metrics like number of Scenario scripts, Key classes, Support classes and Subsystems accumulates the overall system requirements in the Object Oriented scenario [1].

**2.1 Use case Metrics (Scenario Scripts):**

Use case metrics are used to count the number requirements in a scenario [2].

*a. Number of Actors associate with a Use case (NAU)*: This metric count the total numbers of actors associated with a scenario in a Use case diagram. It measures the complexity of a scenario with respect to FP measures. Thus, the number of services requested by end users gives degree of usage of requirements in a scenario.

*b. Number of Messages associated with a Use case (NMU)*: Use case is characterised by set of sequence and collaboration diagrams. This metric is useful in measuring the number of requirements in a scenario provides an input to design phase.

*c. Number of System Classes associated with a Use case (NSCU)*: This metric counts number of system classes whose objects associated with a scenario. It eases the changes to be introduced in the scenario in future.

## 2.2 Requirement quality Metrics:

Use cases and other traditional tools are used to collect and frame the system architecture. However, to check the quality of collected requirements, many requirement quality metrics are discussed in the literature which remains as open issues for further improvement [3].

*a. Unambiguous:*

Requirement collection is a crucial phase of SDLC since ambiguous and wrong requirements may cause system failure. Since analysing a scenario in to classes and objects is up to one's perception, ambiguity may get induced among reviewers leading to an undesired system.

$$QUA = \sum \frac{Ridr}{N} \quad \text{(Eq.1)}$$

Where *Ridr* is number of requirements reviewed identical by reviewers,
*N* is total number of requirements.

*b. Correctness*: A Use case Scenario explores numerous functional requirements. However, the right interpretation of user vocabulary will lead to correct set of valid requirements. Vague requirements like 'shall', 'multi-user', 'user-friendly', are thoroughly analysed as an input to design phase.

$$QC = Nv/(Nnv * N) \quad \text{(Eq.2)}$$

where
*Nv* is number of valid requirements,
*Nnv* is number of still not valid requirements and
*N* is Total number of requirements

*c. Completeness*: It reflects the depth/breadth of requirements in a scenario. Each requirement is unique and need to serve the user as a complete package. A List with Add() and Delete() services is incomplete. Analyst need to have in-depth knowledge of the problem space and identify other modifier and selector services as List properties.

Though many requirement metrics are discussed and practiced, it is hard to quantify the quality of requirements metric. Organisational guidelines/best practices will improvise the requirement collection to analysis process including their traceability and volatility. However, involving the stakeholders in requirement analysis would reduce the confusion and frequent changes due to uncertainty. Having a count on requirement changes other than business change will give an insight on process adapted.

## 3. Software Design Quality Metrics:

There is always a race between traditional System Analysis and Design and recent Object Oriented Analysis and Design methodologies. Service-centric systems with major focus on huge set of services follow traditional technique, whereas Object-Oriented development suits data-centric problem space demanding data security. Design tools like Context diagram through Data Flow Diagram till Decision Tree are enough to express the solution diagrammatically for traditional system, however, in OOAD, UML plays an important role in modelling the architecture with a different set of diagrams. Whichever the methodology adapted, the success of the product is basically relying on quality of the design architecture. There are many such design quality metrics in existence being used to ensure the quality of final product at design level [2].

The solution domain of a problem contains classes, packages or interfaces at different level of complexity. Metrics are categorised from quantity to quality to measure overall quality of a design phase.

### 3.1 Quantity Metrics:

Never the choice of methodology, quantity of components in a design model and number of executable statements are also considered as measure of quality of a design phase.

*a. UML Design Model Metrics*: An UML model represents the design of a solution. It comprises of packages, classes, subclasses, super classes, abstract classes, interfaces and the relationships among them. A count on all, contributes to find complexity of overall design under construction.

*b. Metrics for Methods*: Function/ Method is an operational element of both the methodologies in practice. Quality of a method depends on the concrete logic been written for a given task and thus number of lines of code. Traditional metric LOC has variations with or without blank and comment lines.

### 3.2 Quality Metrics:

Quality of a solution domain is based on quality of each of its components.

*a. Metrics for class:* Class is a basic building block of data-centric system. Plenty of metrics are defined (CK & MOOD) in the literature and various open source and commercial tools are available on these metrics.

**Table 1. Design quality metrics**

| Metric | Formula | Remarks |
|---|---|---|
| **CK METRICS** | | |
| Weighted Methods per Class (WMC) | $WMC(C) = \sum_{i=1}^{n} ci(Mi)$ (Eq. 3) | Measures the complexity of a methods in terms of effort and time for development and maintenance |
| Response For a Class (RFC) | $|RS| = \{M\} \cup$ all I $\{Ri\}$ (Eq.4) where $\{Ri\}$ = set of methods called by method $i$ and $\{M\}$ = set of all methods in the class C | When an object of a class sends a message, the methods executed inside and outside of a class are counted. |
| Number Of Children(NOC) | NOC= No. of level 1 subclasses of a class | Measures the degree of reusability |
| Depth of Inheritance Tree(DIT) | DIT=No. of classes in a hierarchy | Measures vertical growth of Inheritance lattice. |
| Coupling Between Object classes(CBO) | CBO=No. of Services been shared | Measures the interdependency between the classes |
| Lack of Cohesion in Methods(LCOM) | LCOM=No. of pair of methods that do not share attributes | Measures degree of interdependency within a class elements |

Apart from CK metrics, MOOD and Li & Henry worked on existing metrics and redefined a few of them to increase the clarity [4].

**b. Metrics for Package**: Grouping the classes in to a package reduces the complexity of a solution. Forming the package is either based on commonality among the classes with respect to functionality or reusability.

**c. Instability and Abstraction Metric**: The architecture is more stable and extensible when it has more number of abstract classes. Thus, stable package is independent but contains interfaces and abstract classes for further addition. Thus instability is a metric to measure stability of a class or a package.

**d. Instability(I)=$Ce / (Ca + Ce)$**     (Eq.5)
Where
$Ce$ (efferent coupling) of a package is number of outside package classes used by a package classes.
$Ca$ (afferent coupling) of a package number of classes being used by outside package classes.
I= [0,1], where I=1 indicates maximum instable package and I=0 indicates stable package.

**e. Abstractness(A)=$(\sum Ac + \sum I)/N$**     (Eq.6)
Where
$Ac$ is number of abstract classes and $I$ is number of interface in category,
$Nc$ is total number of classes in a category.

**f. Dependency Inversion Principle Metric**: It indicates that the modules in a design either a package or a class must depend on Abstract entities so that it can be easy extendable and modifiable. High-level modules with rich in services should not depend on low-level modules (concrete class) since concrete classes are prone to change periodically as requirement changes in a solution space.

**g. Acyclic Dependency Principle Metric**: The complexity of a system architecture increases when packages in the scenario are more dependent on others. Moreover, when dependency forms a cycle, the architecture becomes too rigid for modification and system would become stagnant in future.

**h. Encapsulation Principle Metric**: A package is highly encapsulated when its sub package is not been much used by lateral or outside packages, but becomes less cohesive when child package/s tightly coupled with outside packages. A good design always supports to divide such packages into two different packages there by achieving 'separation of concern'. EP is 100% when none of the child package is used outside and 0% when all of the children are being used by outside packages.

**4. Implementation Metrics:** This phase is for a coder to exhibit his skills to make the product right as well as user-friendly. Quality metrics are applied at different crucial sub phase of coding to keep the product in line with user requirements [5].

**4.1 Code base Metrics:** Quality of a Code set is not just depends on logic written for a service, but also its availability in future for other projects in different languages with minimum modification.

**a. Testable**: This metric checks whether the code has a logging facility, scriptable interfaces and real-time monitoring capabilities to make the code friendlier to the end user.

*b. Supportable*: This metric checks amount of support given to the users, technical staff, testers, developers by providing enough comments when system goes wrong.

*c. Maintainable*: This metric checks the factors like modularity, reviewability, accessibility so that product maintenance in future maintenance can be easily done with hands on system information.

d. *Portable*: This metric focuses on how easily a product can be deployed on different platform.

### 4.2. Code Coverage Metrics:

Code coverage metrics measures the quality of code written for a task. It checks the relevance of code written and its execution when system is in to operation.

*a.* **Symbol Coverage Metric:** This metric checks the execution of all sequence points in a code set. Sequence points can be nested (loops) and metric checks the code quality for the relevance of such sequence points in every module.

*b.* **Method Coverage Metric:** This metric measures the number of methods has been executed. This metric only tells whether a function is executed, supporting overall project coverage.

*c.* **Branch Coverage Metrics:** This metric focuses on number of branches executed in each module. Each branch represents a block of code and its execution reflects the code coverage.

### 4.3 Code Quality Metrics:
This metric measures the quality level of the code in a product.

### 5. Conclusion:
Quality is an uncompromised factor in software development process. Software deliverables are checked for quality at every phase of development process to ensure the quality of end product. Quality process is organization-specific providing a basement for quality product. Numerous metrics are used by various industries to measure the quality of requirement phase to implementation phase to uphold them in the market. However, research in this area is in continual progress provide better metrics for Product, People, Process ($P^3$) to support the development team. Implementation of apt metrics during the development process ensures production of high quality software there by retaining the total customer satisfaction and improved business in the market.